\documentclass[aps,pre,reprint,amsmath,amssymb,superscriptaddress]{revtex4-2}

\usepackage{graphicx}
\usepackage{dcolumn}
\usepackage{bm}
\usepackage[utf8]{inputenc}

\begin{document}

\title{Impact of Channel Dynamics on Higher-order Interactions of Oscillators}

\author{Gabriel Marghoti}
\email[Corresponding author: ]{gabrielmarghoti@gmail.com}
\affiliation{Departamento de Física, Universidade Federal do Paraná, Curitiba, Paraná, Brazil}
\affiliation{Centro Interdisciplinar de Ciência Tecnologia e Inovação, Universidade Federal do Paraná, Curitiba, Paraná, Brazil}
\author{Isabela A. Martins}
\affiliation{Departamento de Física, Universidade Federal do Paraná, Curitiba, Paraná, Brazil}
\affiliation{Centro Interdisciplinar de Ciência Tecnologia e Inovação, Universidade Federal do Paraná, Curitiba, Paraná, Brazil}
\author{Giovana S. Spezzatto}
\affiliation{Departamento de Física, Universidade Federal do Paraná, Curitiba, Paraná, Brazil}
\affiliation{Centro Interdisciplinar de Ciência Tecnologia e Inovação, Universidade Federal do Paraná, Curitiba, Paraná, Brazil}
\author{Emanuel B. S. A. Cambraia}
\affiliation{Departamento de Física, Universidade Federal do Paraná, Curitiba, Paraná, Brazil}
\affiliation{Centro Interdisciplinar de Ciência Tecnologia e Inovação, Universidade Federal do Paraná, Curitiba, Paraná, Brazil}

\author{Sergio R. Lopes}
\affiliation{Departamento de Física, Universidade Federal do Paraná, Curitiba, Paraná, Brazil}
\affiliation{Centro Interdisciplinar de Ciência Tecnologia e Inovação, Universidade Federal do Paraná, Curitiba, Paraná, Brazil}
\author{Thiago L. Prado}
\affiliation{Departamento de Física, Universidade Federal do Paraná, Curitiba, Paraná, Brazil}
\affiliation{Centro Interdisciplinar de Ciência Tecnologia e Inovação, Universidade Federal do Paraná, Curitiba, Paraná, Brazil}

\date{\today}

\begin{abstract}
Modeling higher-order interactions (HOIs) in nonlinear networks with static topologies is often physically restrictive. We demonstrate that standard 3-body Kuramoto couplings are mathematically equivalent to pairwise connections modulated by latent variables of transmission channels. While standard HOI topologies emerge in the adiabatic limit of these variables, relaxing this constraint reveals that latent channel timescales dictate collective macroscopic states. Specifically, transmission inertia drives bistability for symmetric interaction tensors and anti-phase cluster synchronization for antisymmetric ones. Furthermore, dynamically induced clustering in global topologies emerges as a finite-size effect of the dynamics of the local channels. Ultimately, we show that relying exclusively on static topologies restricts interaction modeling. Integrating latent variables captures the transient inertia and fundamental asymmetry of physical networks, bridging the analytical utility of higher-order functions with the reality of the underlying transmission medium.
\end{abstract}

\maketitle

\section{Introduction}

The synchronization of interacting oscillators is a fundamental emergent property typically analyzed through phase reductions, most notably the Kuramoto model \cite{kuramoto1984chemical}. While the standard formulation assumes strictly pairwise (dyadic) coupling, recent extensions have introduced phenomenological higher-order interactions (HOIs) --- such as the 3-body coupling term $\sin(\theta_j + \theta_k - 2\theta_i)$ --- to capture complex synchronization transitions \cite{bianconi_nature_physics, hoi_standard}. The prevalent reductionist approach maps these terms to static hyperedges, formally represented by an adjacency tensor $B_{ijk}$. However, while higher-order structures offer valuable modeling abstractions, from a fundamental physical perspective, the underlying physical mechanisms are typically rooted in pairwise interactions \cite{peixoto2026graphs}. This perspective is supported by recent formal proofs demonstrating that entire classes of higher-order dynamics on hypergraphs are structurally redundant \cite{Lucas2026, Kirkley2025} and can be exactly reduced to weighted pairwise networks, fully preserving the underlying local dynamics \cite{Llabr_s_2026}. Another example is found in contagious processes, where higher-order dynamics can be closely approximated by pairwise interactions with suitably adjusted parameters \cite{eugene2026}. Furthermore, recent results further highlight the duality between structural and dynamical sources of nonlinearity, demonstrating that nonlinear graph-based dynamics can be reformulated as linear dynamics on higher-order interaction structures \cite{lacasa2026equivalence}.

Recent developments indicate that structural HOIs are often not true mechanistic processes, but rather emergent properties of phenomenological models that substitute complex dynamical mechanisms with effective multi-body coupling terms \cite{Letten2019, calleja2026ecology}. Rigorous phase reduction theories confirm this, demonstrating that multi-body interactions emerge naturally as effective artifacts when expanding pairwise-coupled oscillators beyond the first-order approximation \cite{kuramoto2019, leon2019, nijholt2022emergent}, which can be subsequently mapped onto hyperedges to create effective phase models \cite{bick2024}. These emergent couplings are further corroborated by specific mechanisms, such as uniform time delays in transmission \cite{fujii2026emergencehigherorder}, while even in explicitly multi-body architectures, critical transition dynamics often collapse to an effective pairwise baseline \cite{salazar2026}. Indeed, treating these higher-order couplings as rigid physical topologies frequently leads to dynamical inconsistencies: the model tends to favor states in which the natural dynamics of the system rarely reach \cite{Zhang2024}. This increasingly recognized perspective raises a critical question: if structural HOIs are phenomenological, what underlying fast-timescale transmission mechanisms are being omitted?

Physical systems are intrinsically limited by available transmission channels. Consequently, any network abstraction that bypasses transmission variables is a theoretical simplification whose validity depends heavily on the underlying timescale assumptions. By applying the adiabatic approximation to bypass the dynamics of fast transmission variables, both effective pairwise and higher-order interactions can naturally emerge from more complex underlying dynamical systems \cite{kuehn2026emergent}.

Furthermore, because true interaction rules are typically latent, the decision to model a system via hypergraphs rather than standard graphs must be grounded in principled statistical inference --- a methodological requirement still largely missing from the higher-order network literature \cite{peel2022statistical}. Despite this gap, a growing consensus has emerged suggesting that phenomenological HOIs are strictly necessary to capture complex network dependencies \cite{Lambiotte2019}, yielding significant dynamical consequences. For instance, mixtures of pairwise and HOI terms can selectively shape collective behavior in ways that purely dyadic models cannot capture \cite{Zhang2023}. These multi-body interactions induce synchronization \cite{muolo2025higher,Gambuzza2021}, phase transitions that result in multistability \cite{Skardal2019}, and are now being integrated into general phase reduction frameworks that validate the direct use of hypergraphs when treating phase models \cite{leon2025}. Additionally, higher-order repulsive couplings are shown to drive cluster synchronization on sparse simplicial complexes \cite{ji2025cluster}. In this work, we critically evaluate these phenomenological simplifications, focusing on how higher-order structures mathematically emerge from the adiabatic elimination of underlying transmission dynamics.

Recent literature highlights the role of directed higher-order coupling functions as a mechanism to induce synchronization dynamics \cite{gallo2022synchronization}. Related to this, the symmetry on the coupling tensor can be used as a guiding principle for selecting higher-order coupling functions \cite{leon2026symmetry}. In this work, we demonstrate that the emergent dynamics of the effective reduced model are strictly governed by the symmetry properties of the interaction tensor.

\section{Model}
We propose a generalized Kuramoto model with phases driven by a transmission variable for $N$ oscillators. Similar models used the channels as variables to model adaptive coupling \cite{berner2019multiclusters, seliger2002plasticity, aoki2009co}. Instead of treating the interaction between oscillators $i$ and $j$ as a static structural weight coupled to an instantaneous phase difference, we define a dynamical state variable $u_{ij}(t)$ representing the actual propagated signal (or effective interaction current) from $j$ to $i$:
\begin{equation}
    \dot{\theta}_i = \omega_i +  \frac{1}{N}\sum_{j=1}^N  u_{ij}(t) \sin(\theta_j - \theta_i) \ ,
    \label{eq:theta}
\end{equation}
where $\omega_i$ is drawn from a Gaussian distribution $\mathcal{N}(0,0.1)$ is the natural frequency of oscillators in the mean frequency reference frame.

Crucially, the physical transmission medium --- whether representing neurotransmitter kinetics in spiking neural networks or the peer-group influence in social consensus models --- possesses inherent inertia. To capture this non-equilibrium signal propagation, $u_{ij}(t)$ undergoes a first-order relaxation process characterized by a finite timescale $\tau$, driven directly by the dyadic phase difference and modulated by the local phase environment:
\begin{equation}
 \tau \dot{u}_{ij} = -u_{ij} + K_1 A_{ij} + \frac{K_2}{N} \sum_{k=1}^N B_{ijk}  \cos(\theta_k - \theta_i)\ .
\label{eq:u_dyn}
\end{equation}
Here, $A_{ij}$ denotes the structural pairwise adjacency matrix representing the fundamental dyadic wiring, while $B_{ijk}$ is a phenomenological weighting tensor defining the relevant local neighborhood.

This formulation ensures that the inertia applies to the physical signal itself rather than just an abstract coupling weight. The structural $K_1$ term provides the standard Kuramoto driving force, while the $K_2$ term captures local environmental modulation, reflecting how the instantaneous activity of neighboring nodes affects the limited availability of the transmission medium (e.g., neurotransmitters).

In the context of the emergent higher-order dynamics, the tensor $B_{ijk}$ represents the environmental modulation of the physical interactions. Specifically, it dictates how the state of a third-party oscillator $k$ modulates the physical transmission channel coming from node $j$ into the target $i$. The standard analytical reduction to a static 3-body phase coupling relies strictly on the symmetry condition $B_{ijk} = B_{ikj}$. Physically, this assumes that node $k$ amplifies or suppresses the $j \to i$ connection with the same weight that node $j$ amplifies or suppresses the $k \to i$ connection.

This symmetry condition forces the local mean-field interference to be perfectly isotropic and non-directional around the target $i$. In realistic physical networks, this assumption is highly implausible. Factors such as directed synaptic connections and medium heterogeneities naturally break this reciprocity. Consequently, the widely adopted $(1, 1, -2)$ hyperedge topology is not a universal emergent property, but rather an artificially restricted special case that obscures the fundamental dynamical asymmetries inherent to the system.

 In real-world complex networks, the physical transmission channel $u_{ij}$ does not operate in isolation but is immersed in the local mean-field generated by $B_{ijk} \cos(\theta_k - \theta_i)$. When the surrounding environment is highly synchronized ($\theta_k \approx \theta_i$), phase coherence dynamically amplifies signal propagation to non-local elements. Conversely, a frustrated local neighborhood attenuates the transmitted signal.

\begin{figure}
    \centering
    \includegraphics[width=0.95\linewidth]{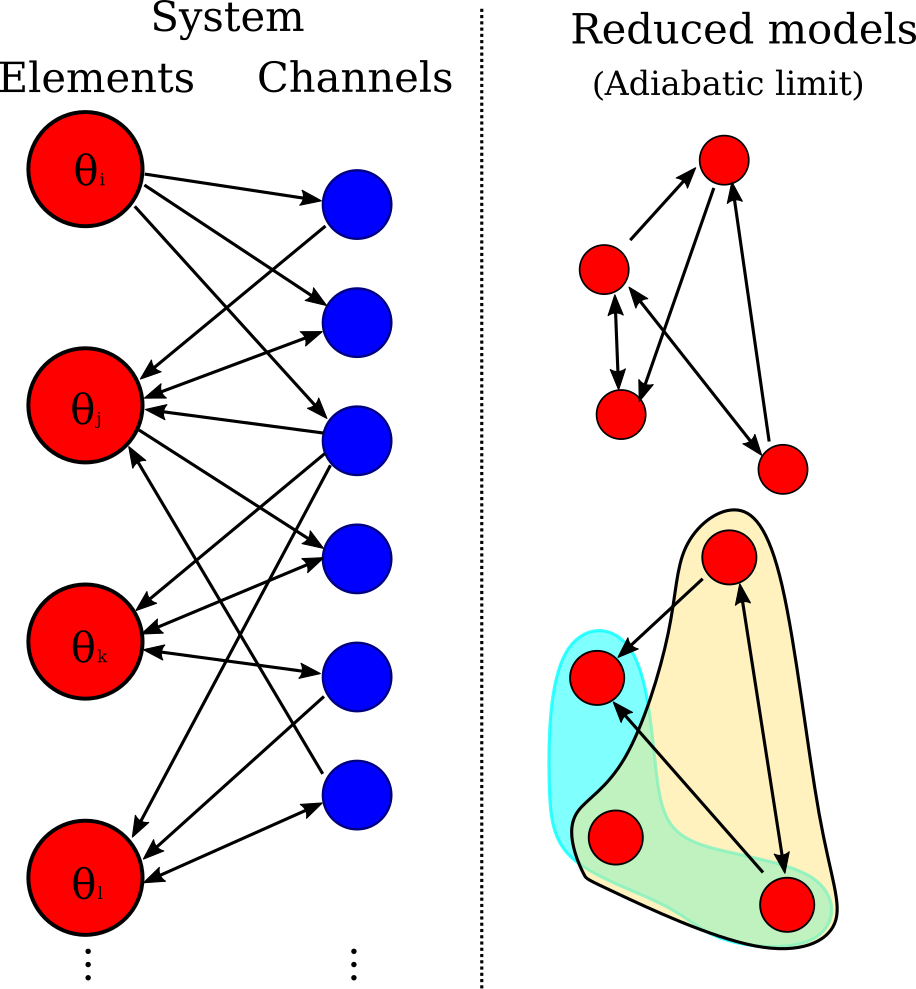}
    \caption{Model reductionism, from physical transmission medium to phenomenological coupling. The left scheme describes the more physically detailed network model, while the right scheme exemplifies possible network reduction based on the underlying dynamics simplification.}
    \label{fig:esquema}
\end{figure}

The latent variable dynamics depends on the phases but also on the coupling parameters and the time scale $\tau$. In specific regimes, the network representation of the system bypasses the $u`s$ dynamics, reducing the representation to a simple and possibly higher-order representation (Fig. \ref{fig:esquema}).

\section{\label{sec:Generalized_Kuramoto_order_parameter}Synchronization Order Parameters}

The Daido order parameters \cite{daido1996multibranch} can be understood as a generalization of the well-known Kuramoto order parameter \cite{kuramoto1984chemical, acebron2005kuramoto}, which is defined by
\begin{equation}
Z_m = \frac{1}{N} \sum_{j=1}^{N} e^{i m \theta_j} ,
\label{eq:daido_order_parameter}
\end{equation}
and quantify the synchronization of the $m$-th harmonic in a system of $N$ coupled oscillators. The modulus $R_m = |Z_m|$ measures the degree of phase coherence of that harmonic. For $m=1$, one recovers the standard Kuramoto order parameter, $R \equiv R_1$, which quantifies global synchronization. A value $R_m = 1$ with $R \approx 0$ indicates that the phases are clustered in $m$ groups, while $R_m = R = 1$ indicates that all phases are perfectly aligned. If $R_m \approx 0$ for all $m$, the phase distribution is incoherent. Higher values of $m$ are essential for detecting multi-cluster states and complex patterns that can arise when the coupling function contains higher-order components.

\section{Dynamical and Structural Limits}

We investigate the theoretical limits of our model by focusing on two distinct reductionist approximations: the time scale of the latent transmission variable $u_{ij}(t)$ (a dynamical reduction) and the symmetry properties of the higher-order interaction tensor (a structural reduction). Both approaches bypass the explicit dynamics of $u_{ij}(t)$, reducing the system to a phenomenological phase-only model governed by $\theta_i$. We demonstrate that taking these specific limits analytically retrieves either the standard pairwise Kuramoto model \cite{kuramoto1984chemical} or the phenomenological higher-order Kuramoto model \cite{bianconi_nature_physics, hoi_standard}.

\subsection{The Standard Pairwise Limit (Static Topology)}

When the local mean-field influence is removed ($K_2 = 0$), the transmission variable acts as a first-order low-pass filter. The exact temporal evolution is given by:

\begin{equation}
    u_{ij}(t) =  u_{ij}(0) e^{-t/\tau} +K_1 A_{ij} \int_0^t \frac{e^{-(t-t')/\tau}}{\tau}  \, dt' \ ,
\end{equation}
with exact solution:

\begin{equation}
    u_{ij}(t) \approx K_1 A_{ij} \left(1 - e^{-t/\tau}\right) + u_{ij}(0) e^{-t/\tau} \ .
\end{equation}
For timescales $t \gg \tau$, the influence of the initial condition $u_{ij}(0)$ and the transient filter delay exponentially decay. The propagated signal functionally converges to $u_{ij}(t) \approx K_1 A_{ij} $. Substituting this into the phase dynamics trivially recovers the instantaneous, purely structural pairwise Kuramoto model \cite{kuramoto1984chemical}.

\subsection{The Adiabatic Limit with Higher-Order Interaction}

When the signal propagation dynamics is rapid compared to the macroscopic phase evolution ($\tau \ll 1$), the transmission variables instantly reach a quasi-steady state. Applying adiabatic elimination ($\dot{u}_{ij} \approx 0$), the fast mediator is dynamically dependent on the macroscopic phases:
\begin{equation}
    u_{ij}(t) \approx K_1 A_{ij} + \frac{K_2}{N} \sum_{k=1}^N B_{ijk} \cos(\theta_k - \theta_i) \ .
\end{equation}
Substituting this instantaneous signals back into the phase equation yields:
\begin{eqnarray}
    \dot{\theta}_i &=& \omega_i + \frac{K_1}{N} \sum_j A_{ij} \sin(\theta_j - \theta_i) \nonumber \\
    && + \frac{K_2}{N^2} \sum_{j,k} B_{ijk} \sin(\theta_j - \theta_i)\cos(\theta_k - \theta_i) \ .
\end{eqnarray}

Assuming symmetric environmental influence ($B_{ijk} = B_{ikj}$), in the adiabatic limit, the model can be reduced to the standard higher-order interaction term used in recent macroscopic modeling, which naturally emerges from a purely pairwise network equipped with fast-timescale transmission dynamics.

In standard phenomenological models, three-body phase interactions are typically introduced via a macroscopic hyperedge tensor $H_{ijk}$, adopting the specific trigonometric form:
\begin{equation}
    \dot{\theta}_i^{(\text{HOI})} \propto \sum_{j,k} H_{ijk} \sin(\theta_j + \theta_k - 2\theta_i) \ .
\end{equation}
However, the dynamic reduction of the transmission variable $u_{ij}$ demonstrates that this standard structural formulation is a restrictive special case rather than a general rule. In the adiabatic limit, the emergent multibody interaction intrinsically couples the phase difference with a local field gradient:
\begin{equation}
    \dot{\theta}_i^{(\text{emergent})} \propto \sum_{j,k} B_{ijk} \sin(\theta_j - \theta_i) \cos(\theta_k - \theta_i) \ .
\end{equation}
This general emergent form does not mathematically map to the standard $(1,1,-2)$ triplet unless the underlying structural tensor exhibits strict permutation symmetry with respect to the environmental sources, $B_{ijk} = B_{ikj}$.

Substituting the quasi-equilibrium solution for $u_{ij}(t) $ into the phase evolution equation generates an emergent product sum for the higher-order interaction ($\Sigma_{\text{HOI}}$):
\begin{equation}
\begin{split}
    \dot{\theta}_i = & \omega_i + \frac{K_1}{N} \sum_j A_{ij} \sin(\theta_j - \theta_i) +\\ & + \frac{K_2}{N^2} \underbrace{\sum_{j,k} B_{ijk} \sin(\theta_j - \theta_i) \cos(\theta_k - \theta_i)}_{\Sigma_{\text{HOI}}} \ .
\end{split}
\end{equation}
To collapse $\Sigma_{\text{HOI}}$, the sum is bisected, and the dummy indices $j$ and $k$ are swapped in the second half:
\begin{equation}
\begin{split}
    \Sigma_{\text{HOI}} = & \frac{1}{2} \sum_{j,k} \Big[ B_{ijk} \sin(\theta_j - \theta_i) \cos(\theta_k - \theta_i) +\\ & + B_{ikj} \sin(\theta_k - \theta_i) \cos(\theta_j - \theta_i) \Big] \ .
\end{split}
\label{eq: sum_HOI}
\end{equation}
Hence, pursuing a clearer physical interpretation of the tensor of environmental influence $B_{ijk}$ and its role in mediating interactions, we can decompose it into its symmetric ($C_{ijk}$) and antisymmetric ($D_{ijk}$) components:
\begin{equation}
    C_{ijk}=\frac{1}{2}(B_{ijk}+B_{ikj})
\end{equation}
\begin{equation}
    D_{ijk}=\frac{1}{2}(B_{ijk}-B_{ikj}) \ .
\end{equation}
By definition, this yields $B_{ijk}=C_{ijk}+D_{ijk}$ and $B_{ikj}=C_{ijk}-D_{ijk}$. Substituing these components back into Eq.(\ref{eq: sum_HOI}) and grouping the terms, we obtain:
\begin{equation}
\begin{split}
    \Sigma_{HOI}= & \frac{1}{2}\sum_{j,k}\Big[C_{ijk}\big(\sin(\theta_j-\theta_i)\cos(\theta_k-\theta_i)+\\ & + \cos(\theta_j-\theta_i)\sin(\theta_k-\theta_i)\big) +\\ & + D_{ijk}\big(\sin(\theta_j-\theta_i)\cos(\theta_k-\theta_i)+ \\ & -\cos(\theta_j-\theta_i)\sin(\theta_k-\theta_i)\big)\Big]
\end{split}
\end{equation}
Applying the trigonometric identities $\sin(x)\cos(y) \pm \cos(x)\sin(y)=\sin(x \pm y)$, we obtain the interaction separated into two distinct physical terms:
\begin{equation}
    \Sigma_{HOI}=\frac{1}{2} \sum_{j,k} \Big[C_{ijk}\sin(\theta_j+\theta_k-2\theta_i) + D_{ijk}\sin(\theta_j-\theta_k)\Big]
\end{equation}
Note that the term $\sin(\theta_j-\theta_k)$ is a nonpairwise interaction in which the nodes $j$ and $k$ jointly influence node $i$ through their phase difference, while the coupling is independent of the state $\theta_i$ of the target node \cite{bick2024}.

From this formulation, we strictly recover the standard 3-body coupling if, and only if, we assume a perfectly symmetric environmental influence ($D_{ijk}=0$, implying full reciprocity where $B_{ijk}=B_{ikj}$). Under this restrictive isotropic condition, the antisymmetric term vanishes, leaving only the purely symmetric hyperedge dynamics:
\begin{equation}
    \Sigma_{\text{HOI}} = \frac{1}{2} \sum_{j,k} C_{ijk} \sin(\theta_j + \theta_k - 2\theta_i) \ .
\label{eq:symmetric}
\end{equation}
This symmetry condition forces the local mean-field interference to be perfectly isotropic and non-directional around the target oscillator $i$. In realistic physical networks, this assumption is highly implausible. Spatial embedding, directed synaptic pathways, and medium heterogeneities naturally break this reciprocity, demonstrating that the standard $(1, 1, -2)$ hyperedge is an artificially restricted special case rather than a general interaction rule.

\subsection{The General case (Finite Transmission Inertia)}

To generalize the strict adiabatic assumption and retain the dynamical information of the transmission channel, we solve Eq.~(\ref{eq:u_dyn}) explicitly. Multiplying by the integrating factor $e^{t/\tau}$ and integrating from an initial state to $t$, the exact state of the mediator becomes a temporal convolution:
\begin{equation}
    \begin{split}
    u_{ij}(t) = & u_{ij}(0) e^{-t/\tau} + \\ & \int_{0}^t  \frac{e^{-(t-t')/\tau}}{\tau} \Big( K_1 A_{ij} + \\ & + \frac{K_2}{N} \sum_{k=1}^N B_{ijk}  \cos(\theta_k(t') - \theta_i(t')) \Big) \, dt' \ .
    \end{split}
\end{equation}

This formulation explicitly exposes the dynamical information lost in static higher-order descriptions. The physical interaction is not an instantaneous $(1,1,-2)$ structural hyperedge, but a function of the past trajectory on the memory kernel function $e^{-(t-t')/\tau} \cos(\theta_k(t') - \theta_i(t'))$. The static 3-body topology is only recovered in the singular limit $\tau \to 0$, where the exponential kernel collapses into a Dirac delta function $\delta(t-t')$, erasing the non-equilibrium inertia of the network.

\section{Results}

To characterize the dynamical regimes of the model, we investigated the dynamical responses of a small network of $N=10$ oscillators with global (all-to-all) coupling, within the ($K_1,K_2$) parameter space under two structural conditions of the environmental tensor $B_{ijk}$: a purely symmetric limit ($B_{ijk}=B_{ikj}$, obtained by setting $C_{ijk}=1.0$ and $D_{ijk}=0.0$ for all indexes), which emulates the perfect reciprocity characteristic of traditional hypergraph approaches in the literature, and an antisymmetric limit ($B_{ijk}=-B_{ikj}$, obtained by setting $C_{ijk}=0.0$ and $D_{ijk}=1.0$ for all entries), which cancels the standard mathematical 3-body interaction (Eq. \ref{eq:symmetric}) in favor of a non-reciprocal flow force.  
Parameter space analysis for larger networks was computationally unfeasible due to the $\mathcal{O}(N^3)$ scaling complexity of the higher-order interactions.

\begin{figure}[ht]
\centering
\includegraphics[width=0.9\linewidth]{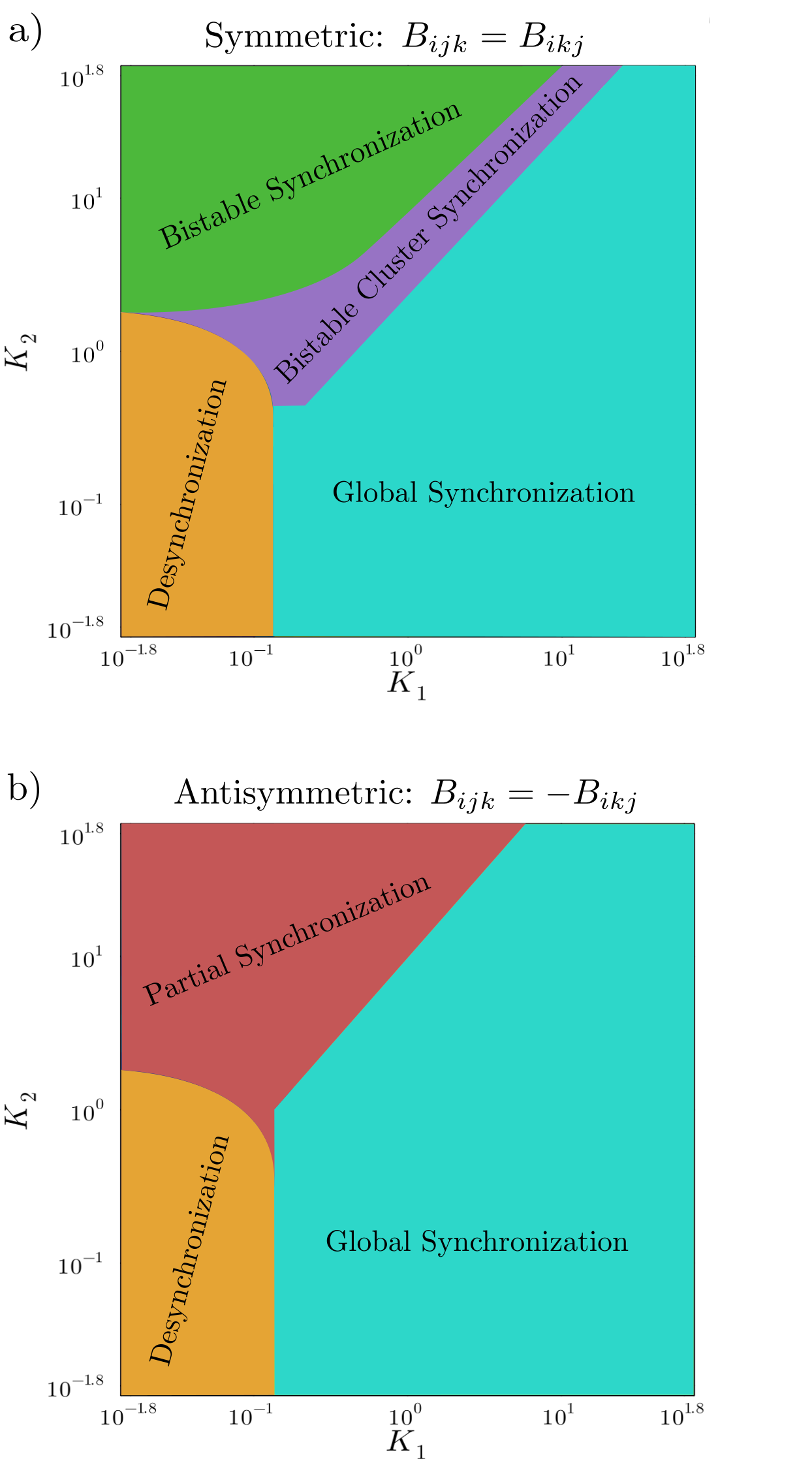}
\caption{Schematic phase diagrams of the synchronization transitions for symmetric ($B_{ijk}=B_{ikj}$) and antisymmetric ($B_{ijk}=-B_{ikj}$) higher-order interactions. The panels illustrate the distinct dynamical regimes as a function of the pairwise coupling strength $K_1$ and the higher-order coupling strength $K_2$ for a 10-oscillator network in the adiabatic limit ($\tau=0.001$). (a) For the symmetric case ($C_{ijk}=1.0, D_{ijk}=0.0$, for all entries excluding the null main diagonal $i=j=k$), the parameter space is partitioned into four regimes: Desynchronization, Global Synchronization, Bistable Cluster Synchronization (coexistence of phase-antiphase synchronized clusters and unsynchronized states), and Bistable Synchronization (coexistence of global synchronized and unsynchronized states). (b) For the strictly antisymmetric case ($C_{ijk}=0.0, D_{ijk}=1.0$, for all entries excluding the null main diagonal $i=j=k$), the parameter space is partitioned into three regimes: Desynchronization, Global Synchronization, and Partial Synchronization.}
\label{fig:phase_diagrams}
\end{figure}

Evaluating the Kuramoto ($R_1$) and Daido ($R_2$, Eq.\ref{eq:daido_order_parameter} for $m=2$) order parameters from both fully synchronized and uniformly spaced initial conditions under an adiabatic fast transmission limit ($\tau=0.001$), we identified five distinct dynamical regimes (Fig. \ref{fig:phase_diagrams}). 

The phase diagrams reveal that symmetric higher-order interactions ($B_{ijk}=B_{ikj}$) fundamentally alter the macroscopic dynamics of the network by inducing strict bistability. Specifically, the Bistable Synchronization regime is characterized by an extreme sensitivity to initial conditions: a synchronized initialization robustly converges to global synchronization ($R_1 \approx 1.0$, $R_2 \approx 1.0$), whereas a uniformly spaced initial phase distribution collapses into complete desynchronization ($R_1 \approx 0.0$, $R_2 \approx 0.0$). Furthermore, strong symmetric interactions can drive the system into Bistable Cluster Synchronization. In this region, depending on the initial state, the network either synchronizes globally or enters into a stable phase-antiphase two-cluster state. This clustered state is rigorously identified by a suppressed primary order parameter ($R_1 < 0.5$) along a maximized secondary order parameter ($R_2 \approx 1.0$), indicating that higher-order terms actively stabilize multi-cluster configurations.

Conversely, introducing antisymmetric higher-order interactions ($B_{ijk}=-B_{ikj}$) suppresses bistability, leading instead to a distinct Partial Synchronization regime. In this dynamically frustrated state, the gradual increase of the pairwise coupling strength $K_1$ acts to progressively recruit oscillators into a synchronized macroscopic cluster. Beyond these topology-specific behaviors, the parameter space for both interaction types is bounded by two universal regimes: absolute Global Synchronization ($R_1 \approx 1.0$, $R_2 \approx 1.0$) and pure Desynchronization ($R_1 \approx 0.0$, $R_2 \approx 0.0$). These fundamental states exhibit invariant behavior within their respective parameter boundaries, dominating the network dynamics regardless of the underlying higher-order interaction symmetry.

\begin{figure*}[ht]
\centering
\includegraphics[width=0.9\linewidth]{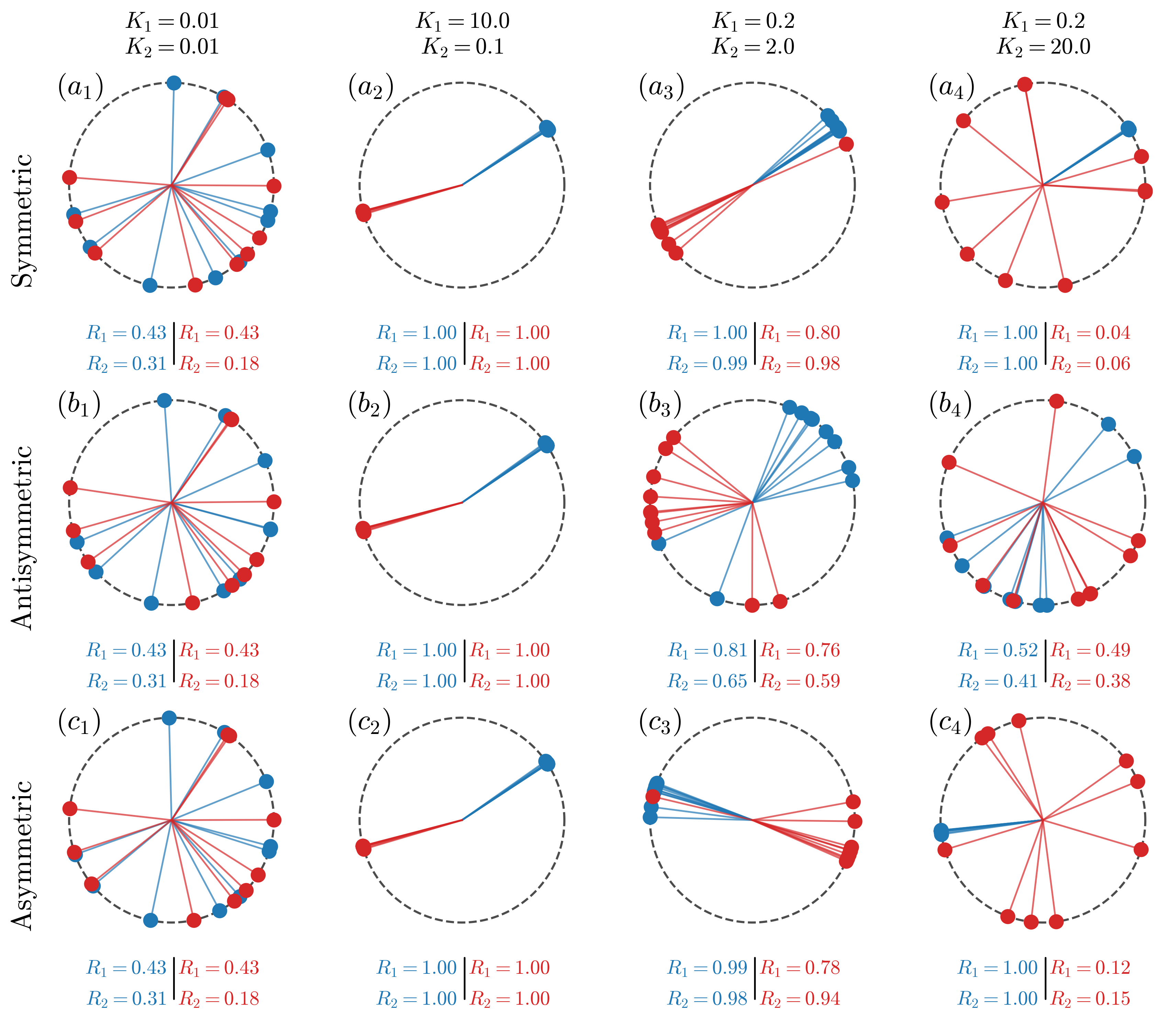}
\caption{Snapshots of the phases of a network of $N=10$ oscillators in the adiabatic regime ($\tau = 0.001$) for representative coupling strengths and network topologies. Colors indicate different initial conditions: red corresponds to uniformly distributed phases, $\theta_i(0)=\frac{2\pi(i-1)}{N}$, whereas blue corresponds to the fully synchronized initial state, $\theta_i(0)=0$ for all oscillators. Panels (a1)–(a4) illustrate the symmetric case, panels (b1)–(b4) the antisymmetric case, and panels (c1)–(c4) the asymmetric case (where the interaction tensor $B_{ijk}$ has unit entries strictly for $j < k$). The coupling strengths $K_1$ and $K_2$ were chosen to represent different regions of the phase diagrams shown in Fig.~\ref{fig:phase_diagrams}.}
\label{fig:phases_snapshots}
\end{figure*}

The phase snapshots in the stationary regime (Fig.~\ref{fig:phases_snapshots}) demonstrate that emergent states depend critically on the competition between pairwise ($K_1$) and higher-order ($K_2$) couplings. Dominant pairwise interactions ($K_1 \gg K_2$, column 2) enforce global synchronization across all topologies, independent of initial conditions. In contrast, strong higher-order interactions ($K_2 > K_1$, columns 3 and 4) promote multi-cluster configurations and partial synchronization. Notably, large $K_2$ induces strict bistability in both the symmetric (a4) and asymmetric (c4) regimes, where the system converges to either fully synchronized or desynchronized states dictated by the initial phase distribution.

\begin{figure}[ht]
\centering
\includegraphics[width=0.95\linewidth]{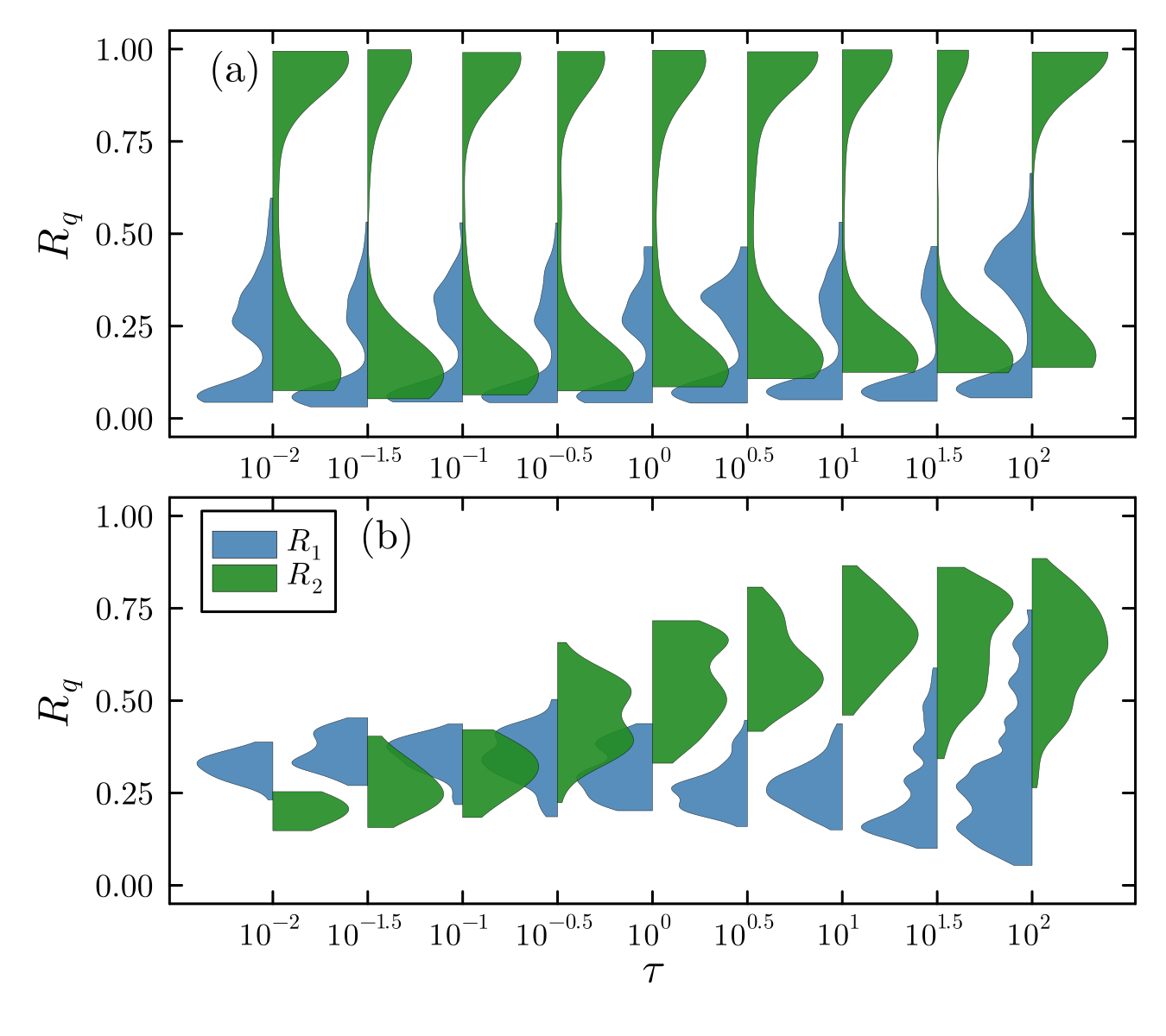}
\caption{Order parameter distributions for (a) symmetric and (b) antisymmetric higher-order interactions as a function of the channel inertia $\tau$. Distributions are computed over the stationary regime of $100$ independent trials in a network of $N=30$, with $K_1 = 0.2$ and $K_2 = 20$. This higher-order dominated regime highlights how intrinsic channel dynamics drive the emergence of two-cluster synchronization in the antisymmetric case, whereas the bistability in the symmetric case remains unaffected.}
\label{fig:tau_sweep}
\end{figure}

Moving beyond the adiabatic limit, we systematically evaluated how the intrinsic timescale $\tau$ of the latent transmission channels influences these macroscopic states. By analyzing the order parameter distributions across varying channel inertia (Fig. \ref{fig:tau_sweep}), we observe a fundamental divergence in the system's behavior. In the higher-order dominated regime ($K_1 = 0.2$, $K_2 = 20$), the symmetric topology reliably maintains bistability regardless of the transmission delay. Conversely, the antisymmetric topology strictly requires non-negligible channel dynamics ($\tau \gtrsim 10^0$) to break partial synchronization and stabilize a two-cluster state ($R_2 \approx 1.0$). This demonstrates that anti-phase cluster synchronization in non-reciprocal networks is not a structural artifact, but a fundamentally dynamic phenomenon driven by the temporal delay in the interaction channels.

\begin{figure}[ht]
\centering
\includegraphics[width=0.95\linewidth]{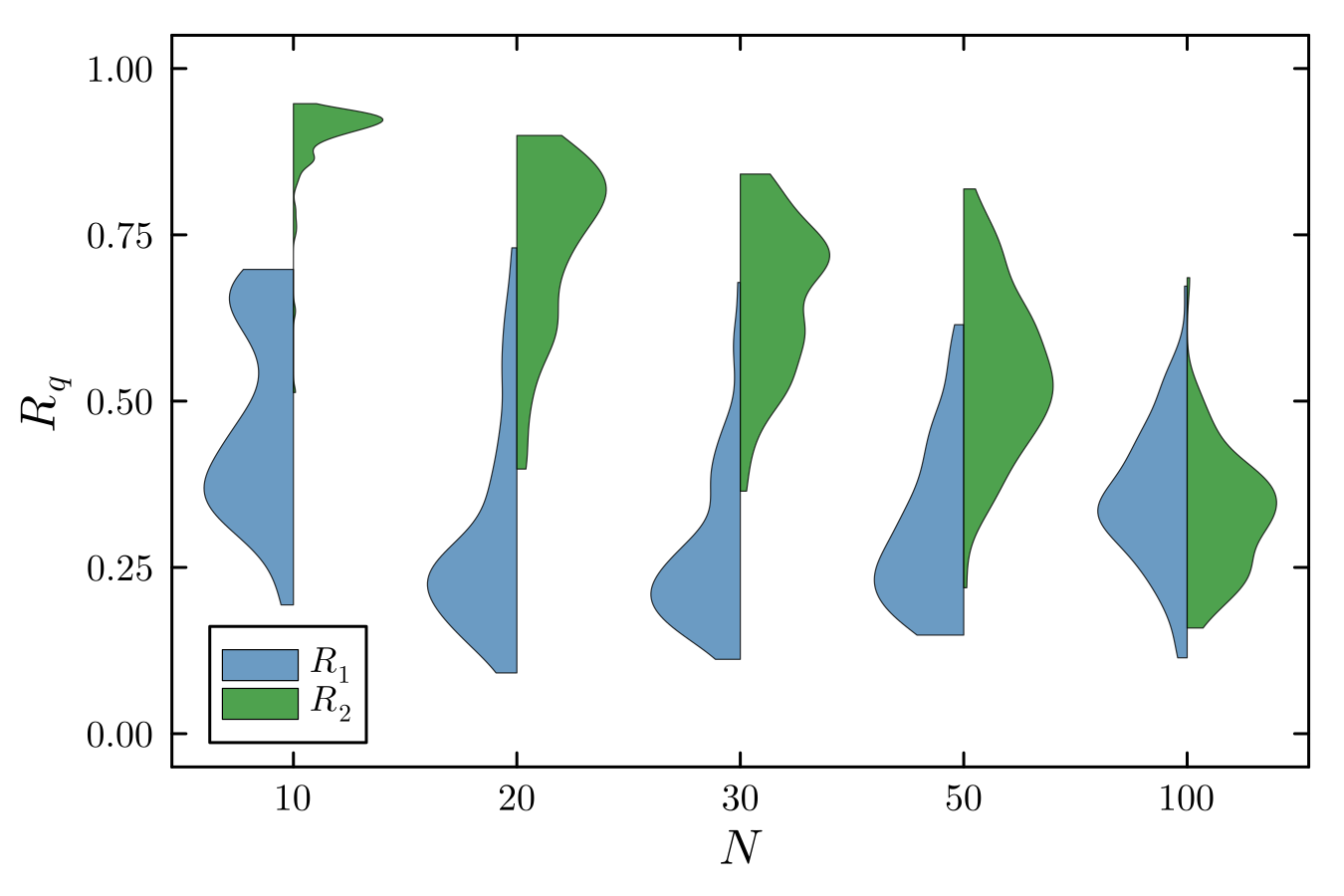}
\caption{Influence of network size $N$ on the cluster synchronization regime observed for antisymmetric higher-order interactions and high inertia channel dynamics, evaluated over 100 independent random trials.}
\label{fig:network_size}
\end{figure}

To determine whether the observed emergent states are fundamental macroscopic properties or finite-size effects, we systematically evaluated the system's behavior across varying network sizes $N$. It is critical to verify if the dynamically induced two-cluster synchronization, unique to the antisymmetric topology, persists in larger ensembles. Figure \ref{fig:network_size} illustrates the distributions of the relevant synchronization order parameters across distinct system sizes, specifically highlighting the slow-transmission regime ($\tau=10$) where the latent channel inertia dominates the network dynamics.

The results reveal that the two-cluster synchronization is highly dependent on the system size (Fig. \ref{fig:network_size}). For small networks, the temporal delay in the interaction channels successfully stabilizes the anti-phase cluster state, evidenced by a Daido order parameter $R_2$ concentrated above $0.5$ while the global Kuramoto order parameter $R_1$ remains consistently suppressed. However, as the network scales to $N \gtrsim 100$, the cluster synchronization deteriorates and $R_2$ vanishes, at least considering trials with random initial conditions. 

Critically, this breakdown indicates that the clustered state in the globally coupled antisymmetric topology is fundamentally a finite-size phenomenon. In dense networks, the $1/N$ normalization constraints (Eq.~\ref{eq:theta} and Eq.~\ref{eq:u_dyn}) dictate that the mean-field dominates as $N \to \infty$. This thermodynamic limit effectively averages out the specific non-reciprocal latent dynamics and structural fluctuations required to separate and sustain the phase clusters. Because the global all-to-all coupling washes out local interactions in large $N$ limits, the dynamic symmetry breaking cannot survive.

The physical mechanism behind this dynamically induced clustering highlights exactly what static hypergraph models overlook. Without transmission inertia, an antisymmetric environmental influence merely drives continuous phase drift. It is the finite memory of the transmission channel that allows delayed, non-reciprocal feedback to actively segregate oscillators into stable, anti-phase populations. However, because this dynamic symmetry breaking relies heavily on localized temporal correlations, the mean-field homogenization of a dense, globally coupled network ultimately overwhelms the delayed feedback as $N$ increases. 

Consequently, to determine if dynamically induced clustering can persist in a true macroscopic limit, future work will focus on analyzing sparse network architectures. We hypothesize that topological sparsity may prevent the mean-field collapse by preserving local heterogeneities and limiting the number of interacting transmission channels, potentially allowing cluster synchronization phenomena to emerge in large-scale systems.

This derivation challenges standard higher-order models, demonstrating that they often mask the true underlying dynamics. While reducing transmission channels to static multi-body interactions offers mathematical utility, enforcing a rigid structural topology artificially restricts the network's behavior. We demonstrated that explicitly modeling the latent transmission variable captures richer dynamical regimes, such as anti-phase clustering, than the restrictive, strictly symmetric functions predominantly used in the literature. However, we critically establish that in dense, globally coupled topologies, these dynamically induced states are ultimately finite-size phenomena that collapse under mean-field dominance as the network scales ($N \gtrsim 100$). Therefore, fully realizing these complex, non-reciprocal dynamics in the thermodynamic limit requires moving beyond all-to-all coupling to explore sparse network architectures explicitly.

\section{Conclusion}

The network abstraction of dynamic interactions fundamentally simplifies complex underlying physical processes, such as those related to transmission medium dynamics. In this work, we demonstrated that standard macroscopic 3-body phase couplings are equivalent to intermediate pairwise connections modulated by a latent transmission variable. By evaluating the network as a variable channel, we show that assigning static hyperedges mathematically erases the non-equilibrium transient dynamics of the physical signal mediator, and at the same time, it limits the interaction function.  

Despite these mechanistic omissions, higher-order structures offer valuable modeling abstractions by simplifying the coupled model. Phenomenological higher-order interactions are often strictly necessary to capture complex network dependencies and shape collective behaviors in ways that purely dyadic models can only capture by extending the modeled variables. These multi-body interactions successfully yield significant dynamical consequences, efficiently modeling emergent phenomena such as synchronization and phase transitions that result in multistability. Pragmatically, substituting complex underlying dynamical mechanisms with effective multi-body coupling terms provides a powerful, computationally efficient reductionist tool when the underlying transmission timescales are negligible.  

However, the mathematical equivalence derived in this work critically challenges the necessity of strictly mapping these effective higher-order terms to structural hypergraphs. In realistic physical scenarios, the influences of local mean-field modulators can be heterogeneous and directed ($B_{ijk} \neq B_{ikj}$). Under these non-reciprocal conditions, the analytical trigonometric terms do not collapse. Consequently, the conventional practice of assigning static (1, 1, -2) hyperedges forces an artificial symmetric constraint onto the network, obscuring a much broader and more complex class of dynamically generated higher-order interactions.  

Therefore, while phenomenological higher-order functions provide vital macroscopic predictive utility, relying exclusively on static topologies artificially restricts interaction modeling. Indeed, recent studies have begun to address this limitation by extending the study of collective dynamics to temporal \cite{anwar2024} and adaptive \cite{PhysRevE.110.064305, kachhvah2022} higher-order structures. A comprehensive theoretical framework must balance the analytical efficiency of higher-order hypergraphs with the physical reality of the underlying transmission medium. To capture the full scope of network dynamics, it remains necessary to include latent variables grounded in pairwise interacting mechanisms, avoiding the exclusion of the fundamental asymmetry and transient inertia inherent to physical networks.

\begin{acknowledgments}
This study was financed in part by the Conselho Nacional de Desenvolvimento Cient\'ifico e Tecnol\'ogico (CNPq), Grants Nos. 168041/2025-1 (GM), 308441/2021-4 (SRL), 407072/2022-5 (SRL), 443575/2024-0 (SRL), 305189/2022-0 (TLP), 408254/2022-0 (TLP), and 158101/2025-1 (IAM); the INCT-NeuroComp (CNPq Grants Nos. 408389/2024-9 and 88887.197665/2025-00); the Coordena\c c\~ao de Aperfei\c coamento de Pessoal de N\'ivel Superior (CAPES) – Brasil (CAPES) – Finance Code 001, Grants Nos. 88887.267448/2026-00 (GSS), 88887.885821/2023-00 (EBSAC), 88881.218671/2025-01 (EBSAC), 88881.895032/2023-01 (SRL), and 88881.189166/2025-01 (TLP); and Funda\c c\~ao Arauc\'aria (Paran\'a State Research Foundation).

\textbf{G. M.:} Conceptualization, Investigation, Visualization, Methodology, Writing - original draft.
\textbf{I. A. M.:} Conceptualization, Investigation, Visualization,  Writing - review \& editing.
\textbf{G. S. S.:} Conceptualization, Investigation, Visualization,  Writing - review \& editing.
\textbf{E. B. S. A. C.:} Investigation.
\textbf{S. R. L.:} Supervision.
\textbf{T. L. P.:} Conceptualization, Supervision, Writing - review \& editing.

\end{acknowledgments}

\section*{Data Availability} % Or Data and Code Availability
The data that support the findings of this article are openly
available \cite{zenodo_marghoti2026_channeldynamics}.

\section*{Declaration of Competing Interest}

The authors declare that they have no conflicts of interest.

\bibliography{references}

\end{document}